\pdfoutput=1
\documentclass[sigplan,10pt]{acmart}

\makeatletter                   %1
\def\mdseries@tt{m}             %1
\makeatother                    %1

\AtBeginDocument{%
  \providecommand\BibTeX{{%
    \normalfont B\kern-0.5em{\scshape i\kern-0.25em b}\kern-0.8em\TeX}}}

\setcopyright{none}
\acmConference{}
\acmYear{}
\acmISBN{} % \acmISBN{978-x-xxxx-xxxx-x/YY/MM}
\acmDOI{} % \acmDOI{10.1145/nnnnnnn.nnnnnnn}
\startPage{1}
\renewcommand\footnotetextcopyrightpermission[1]{}

\newcommand{\systemname}{\sc{Solythesis}}

\newcommand{\sourcecode}[1]{\texttt{#1}}

\newcommand{\rom}[1]{\uppercase\expandafter{\romannumeral #1\relax}}

\usepackage[frozencache=true]{minted}
\usepackage[figure,vlined,linesnumbered]{algorithm2e}
\usepackage{amsmath}
\usepackage{stmaryrd}
\usepackage{pgf-pie}
\usepackage{pgfplots}
\usepackage{esvect}
\usepackage[htt]{hyphenat}
\usepackage{mathpartir}

\graphicspath{{./figures}}
\setminted{
frame=lines,
baselinestretch=0.7,
fontsize=\footnotesize,
linenos,
baselinestretch=1,
breaklines,
xleftmargin=17pt,
escapeinside=??
}

\SetKwInOut{Input}{Input}
\SetKwInOut{Output}{Output}
\SetKw{kwGoTo}{go to}
\SetKw{kwThrow}{throw}
\SetKw{To}{ to }
\SetKw{In}{ in }
\SetKw{Not}{ not }
\SetKw{Continue}{continue}
\SetKw{Is}{ is }
\SetKw{Or}{ or }
\begin{document}
%don't want date printed
\date{}

\title{Securing Smart Contract On The Fly}

\author{Ao Li}
\affiliation {
  \institution{University of Toronto}
}
\email{leo@cs.toronto.edu}
\author{Jemin Andrew Choi}
\affiliation {
  \institution{University of Toronto}
}
\email{choi@cs.toronto.edu}
\author{Fan Long}
\affiliation {
  \institution{University of Toronto}
}
\email{fanl@cs.toronto.edu}

\begin{abstract}

We present {\systemname}, a source to source Solidity compiler which takes a
    smart contract code and a user specified invariant as the input and
    produces an instrumented contract that rejects all transactions that
    violate the invariant. The design of {\systemname} is driven by our
    observation that the consensus protocol and the storage layer are the
    primary and the secondary performance bottlenecks of Ethereum,
    respectively. {\systemname} operates with our novel delta update and delta
    check techniques to minimize the overhead caused by the instrumented
    storage access statements. Our experimental results validate our hypothesis
    that the overhead of runtime validation, which is often too expensive for
    other domains, is in fact negligible for smart contracts. The CPU overhead
    of {\systemname} is only 0.12\% on average for our 23 benchmark contracts.

% We also propose multiple optimization
% techniques that can improve the performance of instrumented smart
% contract significantly.
%{\systemname} employs multiple optimization techniques in order to
%mitigate the overhead introduced by smart contract instrumentation.
%In order to demonstrate the expressiveness of {\systemname} language,
%we specify the specifications for ERC20, ERC721, and ERC1202 standards.
%Our experiment results confirm that the instrumentation of the smart
%contract will not slow down the existing blockchain system. Due to
%the fact that transaction processing
%and smart contract execution is not the bottleneck of the blockchain
%systems, our result highlight the possibilities of bringing runtime
%verifications to smart contracts.

\end{abstract}

\maketitle

\section{Introduction}

Smart contracts are one of the most important features of blockchain
systems~\cite{ethereum}. A smart contract is a program that encodes a set of
transaction rules. Once deployed to a blockchain, its encoded rules are
enforced by all participants of the blockchain network, and therefore it
eliminates counter party risks in sophisticated transactions. People have
applied smart contracts to a wide range of domains such as finance, supply
chain management, and insurance.

Unfortunately, like other programs, smart contracts may contain errors. Errors
inside smart contracts are particularly severe because 1) it is often
impossible or at least difficult to change a smart contract once deployed; 2)
smart contracts often stores and manages critical information such as
digital assets and identities; 3) errors are treated as intended behavior of
the smart contract and faithfully executed by blockchain systems. As a result,
errors inside smart contracts often lead to large financial losses in the real
world~\cite{dao, bec}.

To make smart contracts secure and correct, one approach is to build static
analysis tools. But such static analysis tools are often inaccurate and
generate a large amount of false positives and/or false negatives~\cite{securify}.
Another approach is to formally verify the consistency between the
implementation and specification of a smart contract~\cite{kevm, verisolid, evmstar}. But such
verification processes typically require human intervention and are often too
expensive to apply in practice.

\subsection{Runtime Validation with {\systemname}}

In this paper, we argue that runtime validation is an effective and efficient
approach to secure smart contracts. For traditional programs, with the access
of runtime information, runtime validation techniques can be fully automated
and can typically achieve much higher coverage than static analysis techniques.
The downside of runtime validation is its excessive performance overhead.
However, the Proof-of-Work consensus is the primary performance bottleneck of
existing blockchain systems. For example, the consensus protocol of Ethereum
can only process up to 38 transactions per second, while the execution engine
of Parity~\cite{parity}, a popular efficient Ethereum implementation, can
process more than 700 transactions on an ordinary laptop with a SSD. Therefore,
our hypothesis is that the overhead of runtime validation, which is often too
expensive for other domains, is in fact negligible for smart contracts.

To validate our hypothesis, we design and implement {\systemname}, a novel
runtime validation framework for \break Ethereum smart contracts. Unlike static
analysis and formal verification techniques that attempt to detect errors in
smart contracts offline, {\systemname} works as a source to source Solidity
compiler and detects errors at runtime. {\systemname} provides an expressive
language that includes quantifiers to allow users to specify critical safety
invariants of smart contracts. Taking a potentially insecure smart contract and
the specified invariants as inputs, {\systemname} instruments the Solidity code
of the smart contract with runtime checks to enforce the invariants. The
instrumented contract is guaranteed to nullify all transactions that violate
the specified invariants.

The design of {\systemname} is driven by our observation that the storage layer
is the secondary performance bottleneck of Ethereum after the consensus layer. Our
program counter profiling results show that the execution engine of Parity
spent over 67\% of time on components that are relevant to blockchain state
load and store operations. Such load and store operations are expensive
because 1) it may be amplified to multiple slow disk I/O operations, 2) it is
translated by the Solidity compiler into multiple instructions (up to 11 EVM instructions), and
3) the Solidity compiler uses expensive cryptographic hash functions to compute the
address of accessed state objects. We therefore design the instrumentation
algorithm of {\systemname} to minimize the number of blockchain state accesses.
This enables {\systemname} to generate secure contracts with acceptable overhead
even if Ethereum or future blockchain systems adopt a fast consensus protocol.

One naive approach to enforce the invariant is to just instrument the runtime
checks at the end of each transaction. For sophisticated invariants that
involve iterative sums, maps, and quantifiers, the runtime checks must use
loops to access many blockchain state values, which will be extremely
expensive. To address this challenge, {\systemname} instead uses a novel
combination of \emph{delta update} and \emph{delta check} techniques.
{\systemname} statically analyzes the source code of each contract function to
conservatively determine the set of state values that could be modified and the
set of sub-constraints that could be violated during a transaction. It then
instruments the instructions to maintain these potentially changed values and to
only enforce these potentially violated constraints.

\subsection{Experimental Results}

We evaluate {\systemname} with 23 smart contracts from \break
ERC20~\cite{erc20}, ERC721~\cite{erc721}, and ERC1202~\cite{erc1202} standards.
ERC20 and ERC721 are two Ethereum smart contract standards for fungible and
non-fungible tokens. ERC1202 is a draft standard for a voting system which is
the key process to many blockchain applications. For each standard, we first
compose an invariant and then apply {\systemname} to instrument the smart
contracts.

Our experimental results show that {\systemname} prevents all vulnerable
contracts from violating the defined invariants. The results also validate our
hypothesis --- the instrumentation overhead is negligible with only 0.12\% CPU
usage overhead and 4.2KB/s disk write overhead on average. Our results also
highlight the effectiveness of our instrumentation algorithm. Even in
extreme cases where the consensus protocol is no longer the performance
bottleneck at all, the {\systemname} instrumentation only causes 24\% overhead
in transaction throughput on average. In comparison, if we use the naive
approach that inserts runtime checks at the end of each transaction, the
transaction throughput would be two orders of magnitude smaller. We believe
our results encourage future explorations on new languages, new analyses, and
new virtual machine designs that can further exploit rigorous runtime
validation to secure smart contracts (see
Section~\ref{sec:results:discussion}).

\subsection{Contributions}
This paper makes the following contributions:
\begin{itemize}

  \item \textbf{Runtime Validation for Smart Contracts:} Our experiment results
      show that processing transactions and smart contract execution are not
        the bottleneck of current blockchain systems. We show that runtime
        validation has the potential of significantly increasing the security
        of the smart contracts with only a small or even negligible overhead.

  \item \textbf{{\systemname}:} This paper presents a novel source to source
      Solidity compiler, {\systemname}, which instruments smart contracts
        to reject transactions that violate the specified invariants on the
        fly.

%  \item \textbf{Standard Specifications:} This paper presents
%  an expressive language that allows developers to define standards formally.
%  We further define three standards ERC-20, ERC-721, and ERC-1202 using
%  the language.
  \item \textbf{Instrumentation Optimizations:} This paper presents novel delta
      update and delta check techniques to optimize runtime
        instrumentation.
\end{itemize}

% The remaining paper is organized as follows. Section~\ref{sec:observation}
% presents our observations on the blockchain system performance bottlenecks and
% discusses how it would influence the way we secure smart contracts.
% Section~\ref{sec:example} presents a motivating example of {\systemname}.
% Section~\ref{sec:design} and Section~\ref{sec:impl} present the design and the
% implementation of {\systemname}, respectively. We evaluate {\systemname} at
% Section~\ref{sec:results}. In Section~\ref{sec:related} we discuss related
% work. We finally conclude in Section~\ref{sec:conclusion}.

\section{Observation}
\label{sec:observation}

We next present our observations on Ethereum blockchain performance. Ethereum
uses a modified version of Nakamoto consensus as its consensus
protocol~\cite{Bitcoin, GHOST, ethereum}. It stores transactions in a chain of blocks.
Participants solve proof-of-work (PoW) problems to generate new blocks to
extend the chain.

\noindent \textbf{Consensus Bottleneck:}
Nakamoto consensus and its chain structure limits the performance of Ethereum
--- it generates a new block every 13 seconds and the sizes of the blocks are
limited by its gas mechanism, which measures the size and the complexity
of a transaction~\cite{ethereum}. Currently, each Ethereum block has a gas
limit of 10,000,000~\cite{gas_limit}, and it was 8,000,000
when we performed our experiments.
A simple transaction that only transfers Ether consumes 21,000
gas, and the gas consumption for transactions calling smart contract functions
is higher.

\begin{figure}[t]
  \begin{center}
  \begin{tabular}{ |c|c|c|c|c|c|c|c| }
    \hline
     & \textbf{ERC20} & \textbf{ERC721} & \textbf{ERC1202}\\
     \hline
     \textbf{Native} & 34 & 11 & 9 \\
     \hline
     \textbf{NoConsensus} & 2181 & 1184 & 1439 \\
     \hline
     \textbf{NoConsensus+Empty} & 3647 & 1869 & 2460 \\
     \hline
  \end{tabular}
  \end{center}
\caption{Number of transactions can be processed by Parity client
with different configurations.}
\label{fig:consensus-observation}
\end{figure}

To understand the impact of the consensus layer on the overall performance of
Ethereum, we run experiments with representative transaction traces for the
following contracts: the BEC contract~\cite{bec_contract}, an ERC-20 smart contract for
managing fungible BEC tokens, the DozDoll contract~\cite{doz_contract}, an ERC-721 smart
contract for managing non-fungible DOZ tokens, and an example contract in
the ERC-1202 contract standard~\cite{erc1202}, which is designed for hosting voting on
Ethereum. We use Parity, the most efficient Ethereum client that is publicly available,
to start a private Ethereum network to process these transactions.

Figure~\ref{fig:consensus-observation} presents the number of transactions that
can be processed per second for each contract. Note that we run our experiments
with three different configurations: 1) we use the Ethereum state at the block
height 5,052,259 as the initial state and run PoW consensus to pack and process
transactions (corresponding to the group ``Native'' in
Figure~\ref{fig:consensus-observation}); 2) we remove the PoW consensus limit
so that Parity can process as many transactions as its transaction execution
engine allows (corresponding to the group ``NoConsensus'' in
Figure~\ref{fig:consensus-observation}); 3) we remove the PoW limit and
start Parity with an empty genesis state instead (corresponding to the group
``NoConsensus+Empty'' in Figure~\ref{fig:consensus-observation}).

The results in Figure~\ref{fig:consensus-observation} show that the consensus
protocol of Ethereum is the primary performance bottleneck. Parity only
processes 9 to 34 transactions per second for the ERC-20, ERC-721, and ERC-1202
contracts. If the consensus protocol is disabled, Parity processes 1184 to 2181
transactions per second for the same set of contracts. Therefore, our hypothesis
is that the overhead of runtime validation is negligible for smart contracts,
because the transaction execution engine is not the bottleneck of Ethereum
clients like Parity at all.

%\noindent\fbox{
%    \parbox{0.95\columnwidth}{
%        \noindent \textbf{Observation 1:} The consensus protocol is the primary
%        performance bottleneck of Ethereum clients.
%    }
%}

\noindent \textbf{Storage Bottleneck:} The results in
Figure~\ref{fig:consensus-observation} show that Parity would run much faster
with an empty initial state than with an initial state corresponding to the
real Ethereum network at block 5,052,259. This is because all Ethereum clients,
including Parity, store the blockchain state as a Merkle Patricia Tree
(MPT)~\cite{ethereum} on the disk. Each update on the blockchain
state will be amplified to multiple disk I/O operations depending on the height
of the MPT. When Parity starts with an empty state, the MPT is simpler.
Therefore, there will be fewer I/O operations than starting with
a complicated state.

\begin{figure}[t]
  \begin{center}
  \begin{tabular}{ |c|c|c|c|c|c|c|c| }
    \hline
     & \textbf{Storage} & \textbf{Verifier} & \textbf{EVM} & \textbf{Other}\\
     \hline
     \textbf{ERC20} & 67.0\% & 25.9\% & 3.9\% & 3.2\% \\
     \hline
     \textbf{ERC721} & 73.5\% & 18.3\% & 5.7\% & 2.5\% \\
     \hline
     \textbf{ERC1202} & 73.1\% & 20.5\% & 3.6\% & 2.8\% \\
     \hline
  \end{tabular}
  \end{center}
    \caption{The performance counter results for different
    components of Parity client.}
    \label{fig:pc_counter}
\end{figure}

To better understand the performance impact of the block chain state updates
(i.e., load/store EVM instructions), we profile Parity in our experimental runs
to collect the performance counters of different components in Parity.
Figure~\ref{fig:pc_counter} presents the profiling results. It classifies the
performance counters into four categories: 1) the modules for the blockchain
state updates, including the RocksDB storage layer and the functions in the EVM
interpreter for the load/store Solidity statements; 2) the modules for
verifying signatures in transactions; 3) the modules for other EVM instructions
except loads/stores; 4) all remaining modules.

Our results show that the blockchain state updates account for more than 67\%
of the performance counters for all experimental runs of Parity when we turn
off the consensus layer. The load and store operations to the blockchain state
are particularly expensive because 1) these operations could trigger one or
more disk I/O operations, 2) the Solidity compiler often generates expensive
SHA3 EVM instructions to compute the address operands of these
operations~\cite{solar, verx}, and 3) the Solidity compiler often translates
one state load/store statement into multiple EVM instructions (up to eleven).

%\noindent\fbox{
%    \parbox{0.95\columnwidth}{
%        \noindent \textbf{Observation 2:} The state storage layer is the
%        secondary performance bottleneck of Ethereum clients.
%    }
%}

Therefore, if Ethereum or any other blockchain system could adapt a fast
consensus protocol which eliminates the consensus performance bottleneck, the
state storage layer would become the new bottleneck of the system. This
observation implies that, to reduce the overhead of a runtime instrumentation
tool, one should minimize the number of instrumented load/store statements.

\section{Example}\label{sec:example}

\begin{figure}[t]
  \centering
\begin{minted}{JavaScript}
contract ERC1202Example {
  mapping(uint => uint[]) options;
  mapping(uint => bool) internal isOpen;
  mapping(uint => mapping(address => uint256)) public weights;
  mapping(uint => mapping(uint => uint256)) public weightedVoteCounts;
  mapping(uint => mapping(address => uint)) public  ballots;
  function createIssue(...) public { ... }
  function winningOption(...) public returns (uint option) { ... }
  function vote(uint issueId, uint option) public returns (bool success) {
    require(isOpen[issueId]);
    uint256 weight = weights[issueId][msg.sender];
    // weightedVoteCounts[issueId] [ballots[issueId][msg.sender]] -= weight;
    weightedVoteCounts[issueId][option] += weight;
    // assert(weightedVoteCounts[issueId][option] >= weight);
    ballots[issueId][msg.sender] = option;
    emit OnVote(issueId, msg.sender, option);
    return true;
  }
}
\end{minted}
  \caption{Simplified source code from a voting contract.}
  \label{code:voting}
\end{figure}

We next present a motivating example to illustrate the design of {\systemname}.
Figure~\ref{code:voting} shows a simplified source code from the draft of
ERC1202~\cite{erc1202}, which is a technical standard draft that defines a set
of function interfaces to implement a voting system. A voting system allows a
user to vote different issues, and returns the winning option. This example is
used for illustration purposes in the ERC1202 draft, but it contains a logic
error.

In Figure~\ref{code:voting}, the \sourcecode{vote()} function updates the vote
of a transaction initiator given an option and an issue. The contract
implements \sourcecode{vote()} and other functions with five state variables.
\sourcecode{options} stores the available options of each issue;
\sourcecode{isOpen} stores the current status of the issue;
\sourcecode{weights} stores the weight of each voter on each issue;
\sourcecode{weightedVoteCount} stores the total weighted count of each option on
each issue; \sourcecode{ballots} stores the vote of each voter on each
issue.

In the implementation of the \sourcecode{vote()} function, it first fetches the
weight of the transaction initiator and updates the weighted votes of the given
option and issue (lines 11 and 13 in Figure~\ref{code:voting}). However, the
implementation contains two errors: 1) the original implementation fails to
consider the case where the transaction initiator votes multiple times on the
same issue; 2) it is possible for an attacker to trigger an overflow error at
line 13 to illegally modify the weighted vote count. Note the original
implementation misses line 12 and line 14 which are necessary to avoid these
errors. We next apply
{\systemname} to the contract and describe how {\systemname} instruments the
contract to nullify these errors.

\begin{figure}[t]
  \centering
\begin{minted}{JavaScript}
standard ERC1202 {
  s = Map (a, c) Sum weights[a][b] Over (b) Where ballots[a][b] == c;
  ForAll (x, y) Assert y == 0 || s[x][y] == weightedVoteCounts[x][y];
}
\end{minted}
  \caption{Specification for ERC1202 standard.}
  \label{code:erc1202}
\end{figure}

\noindent \textbf{Specify Invariant:}
Note that both of these errors cause the contract to potentially violate the
ERC-1202 invariant where the total weighted count of an option on an issue
should equal to the sum of all weights of voters who voted for the
option. To apply {\systemname}, we first specify this invariant shown as
Figure~\ref{code:erc1202}. The first line in Figure~\ref{code:erc1202} defines
an intermediate map $s$ that corresponds to the sum of all weights of the
voters that voted for an issue and option pair. The second line defines a
constraint with the \sourcecode{ForAll} quantifier to enforce that for all
pairs of issues and options, the intermediate map $s$ should equal to the
calculated \sourcecode{weightedVoteCount} in the state. Note that in ERC-1202,
zero is a special option id to denote that a voter did not vote yet. Therefore
our invariant excludes the option id zero.

Figure~\ref{code:erc1202} highlights the expressiveness of the invariant
language in {\systemname}. A user can refer to any state variable in the contract
(e.g., \sourcecode{weights}, \sourcecode{ballot}, and
\sourcecode{weightedVoteCount} in Figure~\ref{code:erc1202}), define
intermediate values including maps, and specify constraints on state
variables and defined intermediate values. {\systemname} supports
sophisticated operations such as a conditional sum over values inside maps
(e.g., line 1 in Figure~\ref{code:erc1202}) and a \sourcecode{ForAll} quantifier
to define a group of constraints for multiple state values at once (e.g., line
2 in Figure~\ref{code:erc1202}).

\begin{figure}[t]
  \centering
\begin{minted}{JavaScript}
contract ERC1202Example {
  ?\color{blue}{mapping (uint => mapping(uint => uint)) s;}?
  ?\color{blue}{uint256[] x\_arr;}?
  ?\color{blue}{uint256[] y\_arr;}?
  ...
  function vote (uint issueId, uint option) public returns (bool success) {
    require(isOpen[issueId]);
    uint256 weight = weights[issueId][msg.sender];

    ?\color{blue}{x\_arr.push(issueId);}?
    ?\color{blue}{y\_arr.push(option);}?
    weightedVoteCounts[issueId][option] += weight;

    ?\color{blue}{x\_arr.push(issueId);}?
    ?\color{blue}{y\_arr.push(ballots[issueId][msg.sender]);}?
    ?\color{blue}{assert(s[issueId] [ballots[issueId][msg.sender]] >= weights[issueId][msg.sender]);}?
    ?\color{blue}{s[issueId] [ballots[issueId][msg.sender]] -= weights[issueId][msg.sender];}?
    ballots[issueId][msg.sender] = option;
    ?\color{blue}{x\_arr.push(issueId);}?
    ?\color{blue}{y\_arr.push(ballots[issueId][msg.sender]);}?
    ?\color{blue}{s[issueId] [ballots[issueId][msg.sender]] += weights[issueId][msg.sender];}?
    ?\color{blue}{assert(s[issueId] [ballots[issueId][msg.sender]] >= weights[issueId][msg.sender]);}?

    emit OnVote(issueId, msg.sender, option);
    ?\color{blue}{for (uint256 index = 0; index < x\_arr.length; index += 1) }?
      ?\color{blue}{assert(y\_arr[index] == 0 || s[ x\_arr[index]][y\_arr[index]] == weightedVoteCounts[x\_arr[index]] [y\_arr[index]]);}?
    ?\color{blue}{x\_arr.length = 0;}?
    ?\color{blue}{y\_arr.length = 0;}?
    return true;
  }
  ...
}
\end{minted}
  \caption{Instrumented voting contract.}
  \label{code:vote_secured}
\end{figure}

\noindent \textbf{Instrument the Contract:}
One naive approach to enforce the invariant in Figure~\ref{code:erc1202} is to
instrument a brute force check at the end of every transaction. This would
cause prohibitive overhead because of the iterative sum
operation and the quantifier constraint. It would cost two or even three
nested loops to check the invariant for every transaction. {\systemname}
instead instruments code to perform \emph{delta updates} and \emph{delta
checks} to reduce the overhead. The intuition is to maintain intermediate
values such as $s$ in Figure~\ref{code:erc1202} and to instrument updates and
checks only when necessary. Figure~\ref{code:vote_secured} presents the
instrumented contract code generated by {\systemname} for our example.

\noindent \textbf{Instrument Delta Updates:}
For every intermediate value such as $s$ in Figure~\ref{code:erc1202},
{\systemname} instruments the contract to add it as a state variable. Then for
every write operation to an original state variable in the contract (e.g., line
18 in Figure~\ref{code:vote_secured}), {\systemname} performs static analysis
to determine whether modifying the original variable may cause the
intermediate value to change. If so, {\systemname} instruments code to update
the intermediate value conditionally (e.g., line 17 and 21 in
Figure~\ref{code:vote_secured}).

\sloppypar{
\noindent \textbf{Compute Free Variable Bindings:}
For every state write operation in the contract, {\systemname} computes
\emph{free variable bindings} against each rule in the invariant. For example,
for the write operation to \sourcecode{ballots} at line 18 in
Figure~\ref{code:vote_secured}, {\systemname} determines that the operation may
change the intermediate value defined in the first line of the invariant.
{\systemname} also binds the free variable $a$ in the invariant to
\sourcecode{issueId}, binds $b$ to \sourcecode{msg.sender},
and binds $c$ to \sourcecode{ballot[issueId][msg.sender]}. These bindings indicate that the
write to \sourcecode{ballot[issueId][msg.sender]} may only change the
intermediate value \sourcecode{s[issueId][ballot[issueId][msg.sender]]}. Therefore
{\systemname} instruments code at lines 17 and 21 to only update this value.}

\noindent \textbf{Instrument Delta Check:}
For every state write operation, {\systemname} also checks it against
\sourcecode{Assert} constraints in the invariant. For example, for the write
operation to \sourcecode{ballots} at line 18 in Figure~\ref{code:vote_secured},
{\systemname} first runs its binding analysis against the constraint at line 2
in Figure~\ref{code:voting} which contains a \sourcecode{ForAll} quantifier.
This analysis determines that the write operation may cause the contract to
violate the constraint when $x$ binds to \sourcecode{issueId} and $y$ binds to
\sourcecode{ballot[issueId][msg.sender]}. {\systemname} defines additional
arrays like \sourcecode{x\_arr} and \sourcecode{y\_arr} to collect such free
variable combinations that may lead to constraint violations lines 3 and 4 in Figure~\ref{code:vote_secured}.
{\systemname} then instruments statements at lines 14-15 and 19-20 to
appropriately maintain these arrays. {\systemname} finally instruments a loop
at the end of the transactions to only check these potentially violated
constraints (lines 25-26).

\noindent \textbf{Nullify Errors:}
{\systemname} generates the instrumented program as its output shown as
Figure~\ref{code:vote_secured}. This instrumented program will enforce the
invariant faithfully during runtime and detect any malicious transactions
that cause the contract to violate the invariant. In our example, we deploy the
instrumented contract to Ethereum and intentionally trigger the error by
sending transactions to vote for an issue multiple times. The instrumented
assertion at line 22 catches this error and aborts the offending transactions
as NoOps. Therefore {\systemname} successfully nullifies the error.

\section{Design}
\label{sec:design}

We next formally present the design of {\systemname}. In this section, we use
the notation $s[X/Y]$ to denote replacing every occurrence of $X$ in the
statement $s$ with $Y$. To avoid confusion, we will use ``$\llbracket
\rrbracket$'' instead of ``$[]$'' to denote indexing of map variables. We also
use the notation $\vv{x}$ to denote a list of variables $x_1, x_2, \ldots$.

\subsection{Invariant and Contract Languages}

\begin{figure}
    \input{figures/core-lang-standard.tex}
    \caption{The invariant specification language.}
\label{fig:core}
\end{figure}

\noindent \textbf{Invariant Language:}
Figure \ref{fig:core} presents the syntax of our invariant specification
language with integers, variables, arithmetic expressions, conditional
expressions, intermediate value declarations, and constraints. There are
three types of variables: state variables, intermediate variables, and free
variables. State variables are variables declared in smart contracts and stored
in persistent storage. Intermediate variables correspond to the intermediate
values defined in the invariant rules of the form ``$v=\text{Map}\ldots$''.
Note that in our language we do not distinguish these two kinds of variables,
because {\systemname} instruments code to declare intermediate variables as
state variables. Free variables are only used together with \sourcecode{Map},
\sourcecode{Sum} or \sourcecode{ForAll}. They act as indexes for a defined
intermediate map, as iterators for a sum operation, or as quantifier variables
to define a group of constraints at once.

Expressions are built out of variables and integer constants. ``aop'' represents
an arbitrary binary operator; ``cop'' represents an arbitrary integer
comparison operator. There are three possible types for expressions, integers
for scalar values, maps that have integer keys, and booleans for conditional
expressions. All variables and expressions should be well typed. $e\llbracket x
\rrbracket$ denotes accessing the map $e$ at the index $x$, where $e$ must be
an expression with a map type and $x$ must be a free variable with the integer
type. It is possible to have multi-dimensional maps. In the rest of this
section, the notation $e\llbracket\vv{x}\rrbracket$ is an abbreviation of
$e\llbracket x_1 \rrbracket \llbracket x_2 \rrbracket \llbracket \ldots
\rrbracket$ for accessing such multi-dimensional maps.
%Similar to Solidity and
%EVM, a map is initially empty --- an empty map will map everything to zero or
%another empty map that has one less dimension.

%Free variables are only used in \sourcecode{sum} and \sourcecode{forall} aggregates.
%They are used as placeholders and represent all possible integers.
%The types of free variables are inferred using inference rules described
%in Section~\ref{sec:design:type}. Note if a type of a free variable can not be
%determined statically using type inference rules, the standard is ill-defined.
% \noindent\textbf{Types: }There are two types of types. An elementary type is a type
% of arity 0. {\systemname} supports three elementary types: Integer, Address, and Boolean.
% A mapping type is a type of arity 2, the first one is index type and the
% second is object type.
% A type signature $\Sigma$ is a partial mapping from the
% variables $v$ and $mu$ to $t$.

%\noindent\textbf{Expressions:} Expressions are built out of variables and constants.
%$\oplus$ represents an arbitrary binary operator. An indexing expression is
%represented by $e_1\llbracket e_2\rrbracket$ where $e_1$ is the object expression
%and $e_2$ is the index expression.
%The language considers only well-typed expressions. Types of expressions can be
%determined statically using type inference rules described in
%Section~\ref{sec:design:type}.

The language has two kinds of rules: intermediate value declaration rules and
constraint declaration rules. An intermediate value declaration can define a
map value indexing over a list of free variables and can conditionally and iteratively sum over
state variable values to compute each map entry. For a rule of the form
``$v=$Map $\vv{x}$ Sum $e$ Over $\vv{y}$ Where $c$'', each entry of the intermediate map $v$ is defined as follows:
$$
\forall\vv{x}\; v\llbracket\vv{x}\rrbracket=\sum{_{\vv{y}}}\left\{
    \begin{array}{ll}
        e & \mbox{if } c \mbox{ is True}\\
        0 & \mbox{otherwise}
    \end{array}
\right.
$$
A constraint declaration rule of the form ``ForAll $\vv{x}$ Assert $c$'' checks
all possible assignments of $\vv{x}$ to ensure that $c$ is always satisfied.
Note that free variable lists after \sourcecode{Map} and \sourcecode{ForAll} can be empty.
Therefore, users can use these rules to define scalar values and simple
constraints as well. Also note that free variables in invariants iterate over
all defined keys in maps where those variables are used as indexes. For
example, in ``ForAll $\vv{x}$ Assert $e\llbracket x\rrbracket$'', $x$ should
iterate over values that correspond to defined keys in $e$.

%\subsection{Type Inference}\label{sec:design:type}

%\if 0
%\begin{figure}
%    \input{figures/types.tex}
%    \caption{Type rules for expressions.}
%\label{fig:type}
%\end{figure}
%\fi

%Figure~\ref{fig:type} presents a set of rules defining well-typed
%expressions with respect to a type signature $\Sigma$, which is
%a partial mapping from the variables $v$ and $\mu$ to type $t$.
%There are two types of types. An integer type is an elementary
%type of arity 0. A mapping type is a type of arity 2, the first
%one is the type of index element and the second one is the type of object element.

%Each rule is of the form $\Sigma\vdash exp:t$ which denotes
%that expression $exp$ has the type $t$. All state variables have a data type
%explicitly declared in smart contracts or standard specifications.
%If a variable is used as the index element of a mapping expression with type
%$\text{Mapping}\;i\;t$, the variable has type $i$.
%Both operands of binary expression have integer types.
%The type of \sourcecode{sum} expression is defined recursively by types of $\vv{\mu}$.
%The \sourcecode{sumtype} function shows the algorithm to compute the type
%of \sourcecode{sum} expression. The input of the function is a list of
%types of $\vv{\mu}$. The function creates a mapping type which has n+1 dimension and
%n is the length of $\vv{\mu}$.

\begin{figure}
    \input{figures/core-lang-contract.tex}
    \vspace{-7mm}
    \caption{Core language for smart contracts.}
    \vspace{-3mm}
\label{fig:smart-contract}
\end{figure}
\noindent \textbf{Smart Contract Language:}
Figure~\ref{fig:smart-contract} presents the core language of smart contracts
that we use to illustrate {\systemname}. ``op'' denotes an
arbitrary binary operator. We do not distinguish normal expressions and
conditional expressions in our contract language. There are two kinds of
variables: state variables that may be referred in the invariant and temporary
scalar variables that are local to the contract program.

``load'' and ``store'' are statements for accessing blockchain state variables.
Similar to Solidity, state variables can be either scalar values or maps. ``for
$t_1,t_2,\ldots$ in $v$ \{$s_1$\}'' would iterate over all possible assignment
combinations of $t_1, t_2, \ldots$ based on how $t_1,t_2,\ldots$ are used as
indexing variables for the map $v$ in $s_1$. If any of these variables are not
used as indexing variables for $v$, this statement is undefined. Loop
statements in our language capture the most common usage of loops in Solidity
contracts, and its syntax simplifies the presentation of our instrumentation
algorithm. Note that in Solidity, every state variable has to be declared
before its use. We omit the declaration syntax for brevity.

%The languages of
%constants, variables, and expression are the same as standard specification.
%The statements of a smart contract are similar to statements of C programming
%language including branching, loop, assignment, function call and return.
%Besides the mapping type mentioned in Section~\ref{sec:design:type}, the smart
%contract allows variables with array types: $\mathrm{array}\;t$. Both array
%variables and mapping variables can be indexed in smart contracts.
% A smart contract function
% is a function

% \subsection{Free Variable Binding and Substitution}

\subsection{Instrumentation Algorithm}

\begin{algorithm}[t]
    \SetEndCharOfAlgoLine{}
\Input{Program $P$ as a list of statements and a list of invariant rules $R$}
\Output{The instrumented program as a list of statements}
$P' \leftarrow P$\;
\For {$r \in R$} {
    \If {$r = ``v = \text{Map }\vv{x}\text{ Sum } e \text{ Over }\vv{y} \text{ Where } c;"$} {
        Assume $v$ is fresh. Insert a declaration of $v$ in $P'$.\;
        \For {$s = ``\text{store } a, \_;" \in P$} {
            $\mathcal{B} \leftarrow \mathrm{BindExpr}(a, e) \cup \mathrm{BindExpr}(a, c)$\;
            $b \leftarrow \mathrm{BindCond}(a, c)$\;
            $\mathit{pre} \leftarrow ``\text{if  }c \; \{ \;
                t = \text{load }v\llbracket\vv{x}\rrbracket;$
                $t' = t - e;
                \text{store }v\llbracket\vv{x}\rrbracket, t';\}"$\;
            $\mathit{pre} \leftarrow \mathrm{Rewrite}(\mathit{pre}, \mathcal{B}, b)$\;
            $\mathit{post} \leftarrow ``\text{if  }c \; \{ \;
                t = \text{load }v\llbracket\vv{x}\rrbracket;$
                $t' = t + e;
                \text{store }v\llbracket\vv{x}\rrbracket, t';\}"$\;
            $\mathit{post} \leftarrow \mathrm{Rewrite}(\mathit{post}, \mathcal{B}, b)$\;
            Insert $\mathit{pre}$ before $s$ and insert $\mathit{post}$ after $s$ in $P'$\;
        }
    }
    \ElseIf {$r=``\text{ForAll }\vv{x} \text{ Assert } c;"$} {
        Declare a fresh map $\alpha$ in $P'$ corresponding to $r$\;
        \For {$s = ``\text{store } a, \_" \in P$} {
            $\mathcal{B} \leftarrow \mathrm{BindExpr}(a, c)$\;
            $\mathit{pre} \leftarrow ``\alpha\llbracket \vv{x} \rrbracket = 1"$\; 
            $\mathit{pre} \leftarrow \mathrm{Rewrite}(\mathit{pre}, \mathcal{B}, \langle \bot, \bot \rangle)$\;
            Insert $\mathit{pre}$ before $s$ in $P'$\;
        }
    }
    $P \leftarrow P'$\;
}
\For {$r=``\text{ForAll } \vv{x} \text{ Assert } c;"  \in R$} {
    $\alpha \leftarrow \text{The defined map that corresponds to }r$\;
    $s' \leftarrow ``\text{for } \vv{x} \text{ in } \alpha \; \{ \; \text{assert }c; \; \} \;"$\;
    Insert $s'$ at the end of $P'$\;
}
\Return $P'$\;

    \caption{Instrumentation algorithm.}
\label{fig:instrument}
\end{algorithm}

Figure~\ref{fig:instrument} presents the {\systemname} instrumentation
algorithm. Given a program $P$ as a list of contract statements and an
invariant $R$ as a list of rules, the algorithm produces an instrumented
program $P'$ that enforces the invariant dynamically. The algorithm has two
parts. Lines 2-12 handle the intermediate value declarations in $R$, while
lines 13-24 handle the quantifier constraint rules in $R$.

For every defined intermediate value $v$, the algorithm instruments a fresh
state variable declaration for the value (line 4). The algorithm then inspects
every store statement $s$ in $P$ and instruments the contract to maintain $v$
(lines 5-12). The algorithm first computes possible bindings of free variables
in the definition of $v$ (lines 6-7). A binding is a set of pairs of free
variables in the definition and expressions in the contract. The binding
corresponds to possible entries of $v$ (if $v$ is a map) that $s$ may influence
via state variable write operations. The algorithm prepares statement templates
for updating $v$ (lines 8 and 10), rewrites free variables in these templates
based on the computed bindings (lines 9 and 11), and then instruments the
rewritten statements into $P'$ (line 12). The update strategy is to first
subtract the old expression value (e.g., $\mathit{pre}$ in lines 8-9) before
the execution of $s$, and then add the new expression value (e.g.,
$\mathit{post}$ in lines 10-11) after the execution of $s$. See
Section~\ref{sec:design:binding} for our binding and rewrite algorithms.

For every quantifier constraint rule $r$, the algorithm also instruments the
declaration for a fresh state map variable $\alpha$. Note that because of the
\sourcecode{ForAll} quantifier, $r$ may correspond to multiple constraint instances. To
handle this, the algorithm inspects every store statement $s$ in $P$ and uses
its binding algorithm to determine whether the execution of $s$ may cause some
previously satisfied constraint instance of $r$ to be violated again. If so,
the algorithm sets the corresponding entry in $\alpha$ (lines 16-19) to mark
these instances. The algorithm finally instruments a \sourcecode{for} loop iterating
over $\alpha$ at the end of the contract to check these potentially violated
constraints (lines 21-24).

\subsection{Binding and Rewrite Algorithms}
\label{sec:design:binding}

\begin{figure}[t]
    \input{figures/binding.tex}
    \caption{Binding algorithm.}
\label{fig:binding}
\end{figure}

\noindent\textbf{Free Variable Binding:}
Figure~\ref{fig:binding} presents our binding algorithm. It defines two
functions $\mathrm{BindExpr()}$ and $\mathrm{BindCond()}$ that are used in our
instrumentation algorithm (see Figure~\ref{fig:instrument}). Given a modified
address $a$ in a contract store statement and an expression $e$ in an invariant
rule, $\mathrm{BindExpr}(a, e)$ returns a set of binding maps that maps free
variables in $e$ to indexing expressions in $a$. To compute
$\mathrm{BindExpr}(a, e)$, {\systemname} recursively traverses the structure of
$e$ and looks for the map indexing expressions that match the state variable
in $a$. The fifth rule in Figure~\ref{fig:binding} creates a binding map for
such matching index expressions for the matching free variables.

Note that because $e$ is an expression from the invariant, it can have multiple
instantiations with different free variable assignments. Intuitively, a binding
map corresponds possible free variable assignments for $e$ that the state value
at the address $a$ may influence. Because one free variable may map to multiple
indexing expressions in a binding map, in our notation we represent the binding
map as a set of pairs of free variables and indexing expressions. For example,
$\mathrm{BindExpr}(a,e)=\{\{\langle x_1, e_1\rangle, \langle x_2,
e_2\rangle\}\}$ means that state value changes at the address $a$ may influence
the evaluation of the instantiations of $e$ in which we replace $x_1$ with
$e_1$ and $x_2$ with $e_2$. If $\mathrm{BindExpr}(a, e)$ returns a set that
contains multiple binding maps, it means that the state value changes at $a$
may influence instantiations that are represented by all of these maps.

Given a modified address $a$ and a condition expression $c$ in an invariant
rule, $\mathrm{BindCond(a, c)}$ returns a tuple pair of a free variable and an
expression. The instrumentation algorithm in Figure~\ref{fig:instrument} uses
$\mathrm{BindCond}()$ to handle intermediate value declaration rules only. The
computation of $\mathrm{BindCond()}$ scans for the condition of the form $e_1
== x$, where $x$ is a free variable. If an intermediate value declaration rule
has such a condition, {\systemname} can directly infer that the free variable
$x$ must equal to the expression $e_1$ for all cases.

%Maintaining the aggregates first requires {\systemname} to substitute free
%variables with smart contract expressions. The \sourcecode{bind} function
%described in Figure~\ref{fig:binding} represents the algorithm that maps all
%free variables of a standard statement $s_1$ to smart contract expressions of
%$s_2$. The function iterates over all sub-expressions of $s_1$ and calls
%\sourcecode{bindExpression} function respectively. A free variable $\mu$ is
%mapped to a smart contract expression $e$ if and only if they are index
%expression of both the same dimension and the same state variable, which is
%described in \sourcecode{buildExpression} function.
% Binding occurs between a standard statement $s_1$ and a smart
% contract statement $s_2$.
% The \sourcecode{bind} function stops if $e_1$ is
% not an indexing expression. Otherwise, the \sourcecode{bind} function
% traverses $e_2$ to find the corresponding indexing expression.
% Figure~\ref{fig:state-var} shows the algorithm that extracts the state
% variable used by the indexing expression.
% The mapping results are stored in the map $\psi$.
% Free variables with the same name should be mapped to the same smart contract
% expression and all free variables of an aggregate should be bound to
% after the binding.

\begin{algorithm}[t]
\SetEndCharOfAlgoLine{}
\Input{The original statement $s$, a set of binding maps $\mathcal{B}$ extracted from expressions, and a binding $b$ extracted from conditions.}
\Output{The rewritten statement respecting the bindings.}

$s' \leftarrow \emptyset$\;
\For {$B \in \mathcal{B}$} {
    $s'' \leftarrow s$\;
    $B' \leftarrow B \cup \{b\}$\;
    \For {$x \in \text{FreeVar}, \text{where }$x$\text{ appears in }s$} {
        \If {$x$ appears in $B'$ multiple times as $\langle x, e_1 \rangle, \ldots \langle x, e_k \rangle$} {
            $s'' \leftarrow ``\text{if } \;e_1 == e_2 \wedge \ldots \wedge e_1 == e_k \; \{ s'' \}"$\;
            $B' \leftarrow B' - \{\langle x, e_2 \rangle, \ldots, \langle x, e_k \rangle\}$\;
        }
        \ElseIf {$x$ does not appear in $B'$} {
            Create a fresh variable $t$\;
            Find $v$ in $s$, where $x$ is used as its index\;
            $s'' \leftarrow ``\text{for } \; t \text{ in } v \; \{ s'' \}"$\;
            $B' \leftarrow B' \cup  \{\langle x, t \rangle\}$\;
        }
    }
    \If {$b=\langle x', e'\rangle \land x' \neq \bot$} {
        $s'' \leftarrow s''[x'/e']$\;
        $B' \leftarrow B' - \{b\}$\;
    }
    \For {$\langle x, e \rangle \in B'$} {
        $s'' \leftarrow s''[x/e]$\;
    }
    $s' \leftarrow s' s''$\;
}
\Return $s'$\;

\caption{The definition of $\mathrm{Rewrite}()$.}
\label{fig:rewrite}
\end{algorithm}

\noindent \textbf{Free Variable Rewrite:} Figure~\ref{fig:rewrite} presents the
definition of $\mathrm{Rewrite}()$. Given a statement template $s$ that may
contain free variables, a set of binding maps $\mathcal{B}$ extracted from
expressions with $\mathrm{BindExpr}()$, and a binding tuple $b$ extracted from
conditions with $\mathrm{BindCond}()$, $\mathrm{Rewrite}(s, \mathcal{B}, b)$
produces a new statement with all free variables being rewritten based on the
provided bindings.

The Rewrite algorithm iterates over binding maps in $\mathcal{B}$ and generates
one statement instantiation based on each binding map. For each free variable
in $s$, it detects whether it has exactly one appearance in the binding map. If
the free variable appears multiple times, i.e., the free variable maps to
multiple indexing expressions, the algorithm instruments additional \sourcecode{if}
statement guards to ensure that all of the matched indexing expressions have the
same value (line 7). The algorithm then removes redundant tuples and only keeps
one of these indexing expressions (line 8). If the free variable does not
appear in the binding map, the algorithm would wrap the statement with a
\sourcecode{for} loop to handle this missing binding and add a tuple that maps the
missing free variable to the iterator variable of the loop (lines 10-13).
Therefore, at line 14, the binding map $B'$ should map each free variable in $s$
exactly to each expression. The algorithm then replaces these free variables with
their corresponding expressions (lines 14-18).

\section{Implementation}\label{sec:impl}

We implement {\systemname} for Solidity smart contracts. We use
Antlr~\cite{antlr} and the Solidity parser~\cite{solidity-parser} to parse standard
specifications and Solidity programs. {\systemname} extends the language
described in Section~\ref{sec:design} to support all Solidity features
including contract function calls. 

\subsection{Function Calls}\label{sec:impl:call}

Because state variables are often updated sequentially in a
transaction, and the invariant may be temporarily violated during the middle of
the transaction, {\systemname} should only check the constraints at the end of
each transaction. For function calls, simply inserting those checks
at the end of each function may cause those checks to be triggered multiple
times during a transaction and result in false positives. To this end,
{\systemname} declares an additional global state variable to track the current
function call stack depth of the execution. The instrumented constraint checks
will only execute if it is the entry function of the transaction (i.e., the
stack depth is one).

{\systemname} uses Surya~\cite{surya} to build the call graph of the smart
contract. With the call graph information, {\systemname} can obtain the set of
functions that are reachable from an entry function. For each function,
{\systemname} prunes away instrumented checks for constraints whose status will
never change if the function is the entry function. Specially, smart contracts
in Ethereum can call functions defined in other smart contracts. To guarantee
correctness, {\systemname} over estimates that the inter-contract calls
will call back any function defined inside the contract.

%This implies that {\systemname} needs to create statements to check all
%constraints for functions that has inter-contract function calls. Moreover,
%{\systemname} uses state variables instead of global memory mentioned in
%Section~\ref{sec:impl:global-memory} to store arrays for ``ForAll'' expressions
%because memory are not shared between inter-contract calls.

\subsection{Global Memory}\label{sec:impl:global-memory}

\begin{figure}[t]
\begin{minted}{JavaScript}
uint256 x_slot;
function vote(...) {
  address[] memory x_arr;
  if (call_depth == 0) assembly {
    x_arr := mload(0x40)
    mstore(0x40,add(x_arr,0x280))
    sstore(x_slot,x_arr)
    mstore(x_arr,0x260)
  }
  else assembly {
    x_arr := sload(x_slot)
  }
  ...
}
\end{minted}
\caption{Inline assembly to initialize or load a global array.}
\label{code:global-memory}
\end{figure}

{\systemname} uses global memory arrays to store free variable bindings for
\sourcecode{ForAll} constraint rules (line 10-11, 14-15, and 19-20 in
Figure~\ref{code:vote_secured}). It uses the global memory rather than the
intra-procedure volatile memory or the blockchain state because 1) a
transaction may contain multiple functions and the instrumented code of these
functions all need to access these arrays and 2) accessing global memory is
cheaper than accessing the blockchain state.

Since the global memory array is not natively supported by Solidity,
{\systemname} uses inline assembly to allocate the in-memory array as well as
assigning the start location of the array to an array pointer.
Figure~\ref{code:global-memory} presents the generated global array of
\sourcecode{x\_arr} in Figure~\ref{code:vote_secured}. \sourcecode{x\_slot} is
a state variable stores the start location of \sourcecode{x\_arr} and
\sourcecode{x\_arr} is the array pointer. {\systemname} only initializes
\sourcecode{x\_arr} when the transaction starts and sets the \sourcecode{x\_arr}
to the value of \sourcecode{x\_slot} directly if the call depth is not zero.
{\systemname} further uses \sourcecode{mload} and \sourcecode{mstore}
instructions to load and store data from/to array. 

\subsection{State Variable Caches}\label{sec:variable-cache}

As an optimization, {\systemname} uses stack or volatile memory in Solidity to
cache state variable values. Our second observation in
Section~\ref{sec:observation} indicates that blockchain state load/store
operations are very expensive, involving disk I/O, cryptographic computations, and
multiple EVM instructions. On the other hand, loading/storing values from/to
stack or memory only requires a single instruction.

%The overhead introduced by state variable operations is amplified
%after the smart contract is instrumented because state variables are created
%for both \sourcecode{sum} and \sourcecode{forall} aggregates.

{\systemname} performs static analysis to determine whether the same
blockchain state value is accessed multiple times in a function. For every such
value, {\systemname} creates a temporary variable to cache it. If multiple
functions can access the state value, {\systemname} will create the temporary
variable in the global memory and will use the technique similar to
Section~\ref{sec:impl:global-memory} to enable all functions to access it.
Otherwise, {\systemname} creates the temporary variable in the stack. At the
end of the function, {\systemname} instruments code to write back cached values
to the state variable. This enables the optimized code to only execute one state load
operation and one state store operation for such a state value. Note that the
optimized code is equivalent to the original code because in Ethereum, all
transactions are executed sequentially, i.e., the blockchain state can only be
read/written by a single transaction at one time. 

%State variable operations of dynamic allocated variables such as mapping and
%dynamic array variables are expensive in Solidity smart contracts. To access a
%dynamic allocated mapping variable, Solidity compiler first pushes a global
%slot index and the mapping index of the variable to the stack. It then stores
%the stack values to the memory so that the crypto hash function could load the
%data from the memory and compute the storage address of the variable. The
%latest implementation of Solidity compiler\footnote{v0.5.9} generates 11
%instructions for a single state variable operation. On the other hand,
%loading/storing values from/to stack or memory only requires a single
%instruction. The overhead introduced by state variable operations is amplified
%after the smart contract is instrumented because state variables are created
%for both \sourcecode{sum} and \sourcecode{forall} aggregates.

%In order to mitigate the overhead introduced by state variable operations.
%{\systemname} takes benefits of the design of Ethereum blockchain system,
%that all transactions are executed sequentially, the persistent storage can
%only be read/written by a single transaction at one time. This allows
%{\systemname} to cache state variables to stack and memory and
%state variable operations are redirected to cache variables. {\systemname}
%performs this optimization conservatively and stores cache variables
%back to storage right after operation finishes, which avoids the race condition
%when the state variable is updated by other functions.

\begin{figure}[t]
\begin{minted}{JavaScript}
uint opt_13 = ballots[issueId][msg.sender];
uint256 opt_14 = weights[issueId][msg.sender];
assert(sum_votes[issueId][opt_13] >= opt_14);
x_arr.push(issueId);
y_arr.push(opt_13);
sum_votes[issueId][opt_13] -= opt_14;
opt_13 = option;
x_arr.push(issueId);
y_arr.push(opt_13);
sum_votes[issueId][opt_13] += opt_14;
assert(sum_votes[issueId][opt_13] >= opt_14);
ballots[issueId][msg.sender] = opt_13;
\end{minted}
\caption{Vote function with state variable caches.}
\label{code:var-cache}
\end{figure}

\sloppypar{
Figure~\ref{code:var-cache} presents the code after the state variable cache
optimization of lines 14-21 from \sourcecode{vote} in
Figure~\ref{code:vote_secured}. {\systemname} creates two cache variables
\sourcecode{opt\_13} and \sourcecode{opt\_14} for state variable
\sourcecode{ballots[issueId][msg.sender]} and
\sourcecode{weights[issueId][msg.sender]} (line 1-2). And the operations of
those two state variables are replaced by cache variables. Note that not all
state variables can be cached. For example, {\systemname} does not cache the
state variable \sourcecode{sum\_votes[issueId][opt\_13]} in line 6 because
\sourcecode{opt\_13} is updated in line 7 and the state variable in line 10 is
represents a different state variable. {\systemname} then stores
\sourcecode{opt\_13} back to \sourcecode{ballots[issueId][msg.sender]}.
\sourcecode{opt\_14} is not stored because its value is not updated.
}
% and cache
% variables are stored back to storage after the operations finish.

% The system later stores those cached variables to persistent storage after
% execution finishes.
% Figure~\ref{fig:cache} shows the statements of sum expression
% after cache optimization. $tmp$ is a temporary variables allocated on the stack,
% which stores the value of $v_{\vv{\mu_1}[M]A_2[M]}$. For the rest operations,
% the contract reads/writes the values from/to $tmp$, which avoids accessing
% persistent storage multiple times. {\systemname} creates a temporary variable
% for dynamic allocated storage variable $v$.

\section{Evaluation}\label{sec:results}

We next evaluate {\systemname} on three representative standards:
ERC20~\cite{erc20} for the fungible token standard, ERC721~\cite{erc721} for the
non-fungible token standard, and ERC1202~\cite{erc1202} for the voting standard.
The goal of this evaluation is to measure the overhead introduced by runtime
validation and to understand the effectiveness of the {\systemname}
instrumentation optimizations. All experiments were performed on an AWS EC2
m5.xlarge virtual machine, which has 4 cores and 16GB RAM. We downloaded and
modified Parity v2.6.0~\cite{parity} as the Ethereum client to run our
experiments.
% Since the block generation speed is not determinable in current Ethereum blockchain
% system, we download and modified Parity~\cite{parity}, the most efficient
% Ethereum blockchain system, so that we can pack a specific number of blocks
% at a fixed speed.

\subsection{Methodology}\label{sec:results:methodology}

\noindent\textbf{Benchmark Contracts:} We collected 10 popular ERC20 and ERC721
smart contracts from EtherScan~\cite{etherscan}. To evaluate the effectiveness
of {\systemname}, we also include BecToken (an ERC20 contract) and DozerDoll
(an ERC721 contract), two contracts that we successfully collect their history
transactions from Ethereum. We further include the ERC1202 example we described
in Section~\ref{sec:example} in our benchmark set.

Note that BECToken implements a customized function,
\sourcecode{batchTransfer}, which has an integer overflow error and violates
the total supply invariant specified by ERC-20 standard. This error was
exploited in 2018 and the market cap of BECToken was evaporated in
days~\cite{bec}. The ERC1202 example has vulnerabilities which we described in
Section~\ref{sec:example}. In our experiments, {\systemname} successfully
nullify errors in both of these contracts. We validate that the instrumented
contracts reject our crafted malicious transactions.

\noindent\textbf{Standard Specifications:} We specify invariants for ERC20,
ERC721, and ERC1202 respectively using the language described in
Section~\ref{sec:design}. ERC1202 is a smart contract standard for implementing
a voting contract. The example in the ERC1202 standard draft unfortunately
contains logic errors. See Section~\ref{sec:example} for details.

%The existing standard only specifies a list of
%interfaces that should be implemented, and defines a few constraints verbally.
%For example, the ERC1202 standard requires that for all issues, the weighted vote
%counts of each option should equal to the sum of weights of voters who voted for that
%option.  Since, the weight of each voter, vote of each voter, and weighted vote
%counts of each option are all stored in state variables in the example voting
%contract, we further look into the implementation of the smart contract to find
%corresponding state variables.

\begin{figure}[t]
\begin{minted}{JavaScript}
standard ERC20 {
  sum_balance = Map () Sum balances[a] Over (a) Where true;
  ForAll () Assert totalSupply == sum_balance;
}
\end{minted}
\caption{Standard for ERC20.}
\label{code:erc20}
\end{figure}
\begin{figure}[t]
\begin{minted}{JavaScript}
standard ERC721 {
  sum_tokenCount = Map () Sum _ownedTokensCount[a] Over (a) Where a != 0;
  ForAll () Assert sum_tokenCount == _allTokens.length;
  sum_ownersToken = Map (b) Sum 1 Over (a) Where _tokenOwner[a]==b && _tokenOwner[a] != 0;
  ForAll (a) Assert _ownedTokensCount[a] == sum_ownersToken[a];
}
\end{minted}
\caption{Standard for ERC721.}
\label{code:erc721}
\end{figure}

\noindent\textbf{ERC20:}
ERC20 is an important smart contract standard that defines the contract
interface and specification of implementing fungible digital assets.
Figure~\ref{code:erc20} shows the invariant for ERC20 standard.
\sourcecode{balances} and \sourcecode{totalSupply} are two state variables in
BECToken that store the balance of each address and the total supply of the
token. The invariant specifies that the sum of account balances is equal to
total supply.

\noindent\textbf{ERC721:} Similar to ERC20, ERC721 is a smart contract standard
of implementing non-fungible digital assets. Figure~\ref{code:erc721} shows
our invariant for the ERC721 standard. \sourcecode{sum\_tokenCount} and
\sourcecode{sum\_ownersToken} are two intermediate variables created by the invariant
to track the state of the contract. \sourcecode{sum\_tokenCount} stores the
number of minted tokens, which is the number of tokens whose owner is not 0, and
\sourcecode{sum\_ownersToken} stores the number of tokens owned by each
address. The invariant specifies that \sourcecode{sum\_tokenCount} equals to the
length of \sourcecode{\_allTokens}, and the number of tokens owned by each user
equals to the values stored in \sourcecode{\_ownedTokensCount}.
\sourcecode{\_allTokens} and \sourcecode{\_ownedTokensCount} are two state
variables declared in ERC721 smart contracts.

%\noindent\textbf{ERC1202:} ERC1202
%is a smart contract standard for implementing a voting contract.
%See Section~\ref{sec:example} for details.
%Figure~\ref{code:erc1202} shows constraints for ERC1202 standard, which
%specifies that the sum of votes for each option equals to the sum of weights of
%users whose vote is that option. ERC1202Example is an example voting contract
%provided by the authors of ERC1202. As described in Section~\ref{sec:example},
%there is an implementation error in the \sourcecode{vote} function which allows
%a user to vote the option multiple times and the {\systemname} protects the
%voting contract from such attacks by adding runtime checks.

\noindent\textbf{Benchmark Trace Generation:}
We crawled the Ethereum blockchain and collected the real transaction history
of BecToken and DozerDoll. We chose BecToken because its history contains
attacks. We chose DozerDoll because it is an ERC721 token, and it has a long
transaction history for our experiments. Note that the collected history
transactions depend on the blockchain state (e.g., the token balance of
accounts), so we cannot reproduce them directly. To address this issue, we
first create a mapping that maps real world addresses to local addresses that
are managed by the Parity client. For each transaction, we replaced the
addresses of the transaction sender and receiver, as well as addresses in
transaction data.

For the remaining contracts, we developed a script to automatically generate random
transaction traces that exercise core functionalities of ERC20, ERC721, and
ERC1202. For ERC20 and ERC721, the generated trace contains mainly
\sourcecode{transfer} transactions. For ERC1202, the generated trace contains
transactions that call \sourcecode{createIssue} and \sourcecode{vote} functions
repeatedly. Each \sourcecode{createIssue} transaction is accompanied by five
\sourcecode{vote} transactions created by different voters respectively.

\noindent\textbf{Experiments with PoW Consensus:}
We apply {\systemname} to instrument all of the 23 benchmark contracts. We then
run Parity 2.6.0 to start a single node Ethereum network to measure the
overhead of instrumented contracts. For BecToken and DozerToken, we use the
collected Ethereum history trace. For the ERC1202 example contract and other
contracts, we use the generated trace. To run a contract on a transaction
trace, we initialize the network with the first 5,052,259 blocks downloaded
from Ethereum main net. We then feed the transactions in the trace into the
network.

In this experiment, we deploy the same smart contract to the blockchain
multiple times and take the average results. To address the randomness of the
PoW consensus process, we modified Parity client so that Parity generates new
blocks at a fixed speed of 1 block per ten seconds, which is the generation
speed upper bound Ethereum ever reaches with its difficulty adjustment
mechanism. Note the gas usage for each transaction is set to the gas usage of
executing the original smart contract, which allows Parity to pack the same
number of transactions for both the instrumented and the original smart
contract. We set the block gas limit to 8,000,000. We monitor the CPU usage and disk
IOs of the Parity client for 500 blocks (\textasciitilde 1.4 hours). Because
the consensus is the performance bottleneck and there is no throughput
difference between the original contracts and the instrumented contracts, we
measure the resource consumption as the instrumentation overhead in this set of
experiments.

%In this experiment, we deploy the same smart contract to the blockchain
%multiple times and take the average results.
%For each transaction, we duplicate the transaction
%and send different transactions to different smart contracts until the sum of
%gas usage of those transactions exceeds 10,000,000, which is the block
%gas limit of the current Ethereum main net~\cite{gas_limit}.
%Note the gas usage for each transaction is set to the gas usage of executing
%the original smart contract, which allows Parity to pack the same number of
%transactions for both the instrumented and the original smart contract.

%Next,
%we submit those transactions to the Parity client, where the transactions are
%executed, packed into a block, and stored in persistent storage in succession.
%We also modified Parity client so that Parity generates new blocks at a fixed
%speed of 1 block per ten seconds. We monitor the CPU usage and disk IOs of the
%Parity client for 500 blocks (\textasciitilde 1.4 hours).

\noindent\textbf{Experiments without PoW Consensus:}
We modified the Parity client to remove the proof of work consensus, but
preserved all required computations and storage operations for the transaction
execution. To evaluate the potential overhead of {\systemname} when the
consensus is no longer the performance bottleneck, we apply {\systemname} to
instrument all benchmark contracts and run the modified Parity client to
measure the overhead. To evaluate the {\systemname} instrumentation
optimizations, we also run naively instrumented contracts that iteratively
check invariants at the end of transactions for comparison. Specifically, we
compare the transaction throughput of the original contracts and the
instrumented contracts. Note that similar to the previous set of experiments,
we initialize the Parity client with the first 5,052,259 blocks from the
Ethereum main net.

\noindent\textbf{\textbf{}}

%\noindent\textbf{Blockchain Configuration:}
%We use {\systemname} to generate both optimized
%and non-optimized instrumented smart contracts in order to measure the
%effectiveness of optimizations.
%We modified Parity client
%which preserves all required computations and storage operations but
%removes consensus verifications as well as proof of works. This allows
%us to monitor the performance of Parity client of processing arbitrary
%transactions. Besides, experiments are carried under two different
%blockchain settings:
%\begin{itemize}
%    \item \textbf{Empty} is an empty blockchain and only contains the
%    genesis block.
%    \item \textbf{Main Net} contains the first  5,052,259 blocks
%    downloaded from Ethereum main net.
%\end{itemize}

\begin{figure*}[t]
    \begin{center}
    \begin{tabular}{ |c|c|c|c|c|c|c|c|c|c|c|c|c| }
    \hline
    & & \textbf{BEC} & \textbf{USDT} & \textbf{ZRX} & \textbf{THETA} & \textbf{INB}
    & \textbf{HEDG} & \textbf{DAI} & \textbf{EKT} & \textbf{XIN} & \textbf{HOT} & \textbf{SWP} \\
    \hline
    \textbf{CPU}& \textbf{S}& 3.11& 3.31& 3.12& 3.11& 3.04& 3.34& 3.13& 3.13& 3.09& 3.29& 3.32 \\
    \cline{2-13}
    \textbf{(\%)}& \textbf{O}& 3.11& 2.99& 3.02& 3.10& 3.02& 3.14& 3.09& 2.92& 3.08& 3.04& 3.12 \\
    \hline
    \textbf{Disk}& \textbf{S}& 96.8 & 97.4& 96.4& 96.3& 96.7& 97.0& 96.5& 96.6& 96.2& 96.6& 96.7 \\
    \cline{2-13}
    \textbf{(KB/s)}& \textbf{O}& 96.6 & 96.6& 96.1& 96.2& 96.5& 96.8& 96.3& 96.1& 96.1& 96.2& 96.6 \\
    % \textbf{Average CPU}& \textbf{S} & 0.912\% & 1.642\% & 1.491\%
    % & 1.423\%& 0.818\%& 0.952\%\\
    % \cline{2-8}
    % \textbf{Usage}& \textbf{S-N} & 1.072\% & 1.642\% & 1.656\%
    % & 1.525\%& 0.850\%& 1.013\%\\
    % \cline{2-8}
    % & \textbf{O} & 0.896\% & 1.642\% & 1.457\%
    % & 1.319\%& 0.799\%& 0.933\%\\
    \hline
    \end{tabular}
    \end{center}
    \begin{center}
    \begin{tabular}{ |c|c|c|c|c|c|c|c|c|c|c|c|c|c| }
    \hline
    & & \textbf{DOZ} & \textbf{MCHH} & \textbf{CC} & \textbf{CLV} & \textbf{LAND} & \textbf{CARDS}
     & \textbf{KB} & \textbf{TRINK} & \textbf{BKC} & \textbf{PACKS} & \textbf{EGG} & \textbf{VOTE}\\
    \hline
    \textbf{CPU}& \textbf{S}& 1.97& 1.99& 1.98& 1.95& 1.89& 1.92& 1.89& 1.90& 1.94& 1.93& 2.04& 2.81 \\
    \cline{2-14}
    \textbf{(\%)}& \textbf{O}& 1.83& 1.80& 1.91& 1.94& 1.81& 1.77& 1.83& 1.79& 1.91& 1.86& 1.82& 2.51 \\
    \hline
    \textbf{Disk}& \textbf{S}& 85.7 & 77.5& 73.9& 77.9& 77.7& 76.2& 75.9& 76.9& 80.6& 74.6& 76.4& 83.4 \\
    \cline{2-14}
    \textbf{(KB/s)}& \textbf{O}& 74.5 & 68.2& 72.2& 69.5& 68.8& 69.5& 66.6& 67.9& 75.3& 70.5& 69.6& 71.1 \\
    \hline
    \end{tabular}
    \end{center}
    \caption{Resources usage for Parity client. ``S'' corresponds to {\systemname} results. ``O'' corresponds to original contract result.}
    \label{fig:cpu}
\end{figure*}

\pgfplotstableread[row sep=\\,col sep=&]{
        interval & carT & carD & carR \\
        BEC      & 1.24   & 68.3 & 68.1 \\
        USDT     & 1.86   & 78.9 & 75.9 \\
        ZRX      & 0.986  & 77.9 & 77.4 \\
        THETA    & 0.967  & 78.4 & 79.2 \\
        INB      & 1.24   & 74.6 & 74.8 \\
        HEDG     & 1.21   & 63.2 & 64.3 \\
        DAI      & 1.11   & 64.0 & 64.9 \\
        EKT      & 1.03   & 68.0 & 66.1 \\
        XIN      & 0.984  & 73.9 & 74.3 \\
        HOT      & 1.03   & 69.1 & 70.7 \\
        SWP      & 1.04   & 67.5 & 67.1 \\
        }\erctwenty

\pgfplotstableread[row sep=\\,col sep=&]{
        interval & carT & carD & carR \\
        DOZ      & 14.6   & 76.4 & 90.1 \\
        MCHH     & 8.37   & 71.5 & 79.0 \\
        CC       & 13.3   & 67.1 & 83.6 \\
        CLV      & 9.16   & 77.4 & 84.5 \\
        LAND     & 7.8    & 66.5 & 80.6 \\
        CARDS    & 7.56   & 74.2 & 78.6 \\
        KB       & 8.26   & 71.7 & 76.3 \\
        TRINK    & 8.21   & 69.9 & 80.9 \\
        BKC      & 7.93   & 69.5 & 77.6 \\
        PACKS    & 8.73   & 72.0 & 83.4 \\
        EGG      & 8.14   & 71.5 & 81.6 \\
        VOTE     & 1.67   & 47.0 & 78.7 \\
        }\ercseven

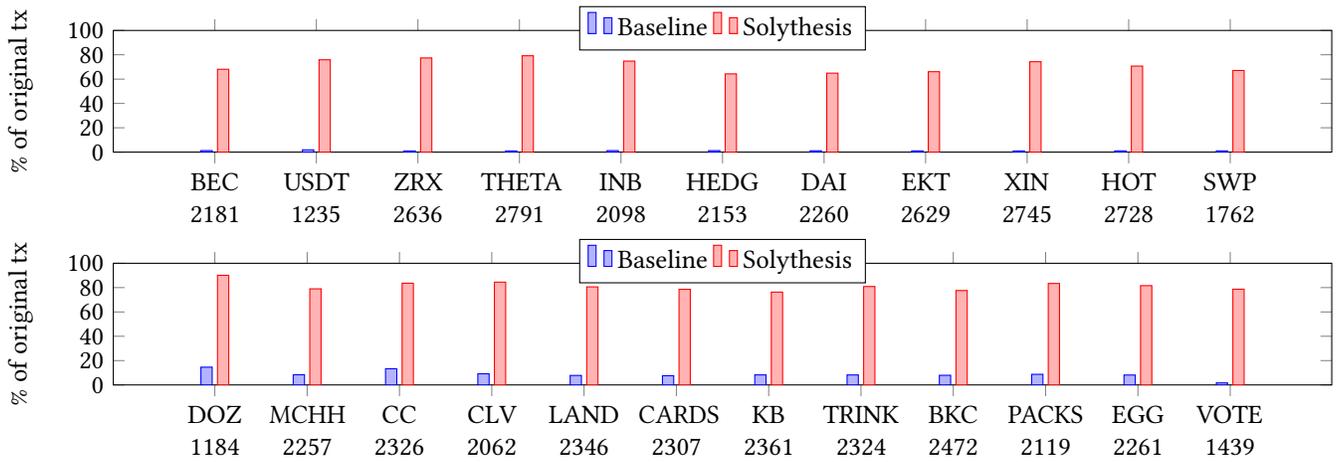
\begin{figure*}[t]
    \begin{tikzpicture}
    \begin{axis}[
            ybar,
            bar width=.15cm,
            width=\textwidth,
            height=.18\textwidth,
            x tick label style  = {text width=1cm,align=center},
            legend style={at={(0.5,1.2)},
                anchor=north,legend columns=-1},
            symbolic x coords={BEC,USDT,ZRX,THETA,INB,HEDG,DAI,EKT,XIN,HOT,SWP},
            xticklabels={BEC 2181,USDT 1235,ZRX 2636,THETA 2791,INB 2098,HEDG 2153,DAI 2260,EKT 2629,XIN 2745,HOT 2728,SWP 1762},
            xtick=data,
            %nodes near coords,
            %nodes near coords align={vertical},
            ymin=0,ymax=100,
            ylabel={\% of original tx},
        ]
        \addplot table[x=interval,y=carT]{\erctwenty};
        %\addplot table[x=interval,y=carD]{\erctwenty};
        \addplot table[x=interval,y=carR]{\erctwenty};
        \legend{Baseline, Solythesis}
    \end{axis}
    \end{tikzpicture}
    \begin{tikzpicture}
    \begin{axis}[
            ybar,
            bar width=.15cm,
            width=\textwidth,
            height=.18\textwidth,
            x tick label style  = {text width=1cm,align=center},
            legend style={at={(0.5,1.2)},
                anchor=north,legend columns=-1},
            symbolic x coords={DOZ,MCHH,CC,CLV,LAND,CARDS,KB,TRINK,BKC,PACKS,EGG,VOTE},
            xticklabels={DOZ 1184,MCHH 2257,CC 2326,CLV 2062,LAND 2346,CARDS 2307,KB 2361,TRINK 2324,BKC 2472,PACKS 2119,EGG 2261,VOTE 1439},
            xtick=data,
            %nodes near coords,
            %nodes near coords align={vertical},
            ymin=0,ymax=100,
            ylabel={\% of original tx},
        ]
        \addplot table[x=interval,y=carT]{\ercseven};
        %\addplot table[x=interval,y=carD]{\ercseven};
        \addplot table[x=interval,y=carR]{\ercseven};
        \legend{Baseline, Solythesis}
    \end{axis}
    \end{tikzpicture}
    \caption{Overhead Comparison with respect to original contract for top ERC20, ERC721 and ERC1202 contracts.}
    \label{fig:baseline-ERC}
\end{figure*}

% \begin{figure*}[t]
%     \begin{center}
%     \begin{tabular}{ |c|c|c|c|c| }
%     \hline
%      & \textbf{SSD-12500} & \textbf{SSD-300}& \textbf{SSD-750} & \textbf{HDD}\\
%      micros/op & \textbf{SSD-12500} & 68.098& 80.940\\
%      MB/s& \textbf{SSD-12500} &  3.6 & 3.8\\
%     \hline
%     \end{tabular}
%     \end{center}
%     \caption{Number of transactions can be processed by parity client per second.}
%     \label{fig:tps}
% \end{figure*}

\subsection{Results with Consensus}

%Figure~\ref{fig:cpu} and Figure~\ref{fig:tps} show the experiment result.
%There is a column in the table for each benchmark.
% For each experiment, we measure the performance of instrumented smart contracts
% (\textbf{S}), instrumented smart contracts without optimization (\textbf{S-N}),
% and the origin smart contract (\textbf{O}) respectively.

% The first column describes the name of benchmark
% and for the rest columns, each column represents a type of smart contract.
% \textbf{Secured} are smart contracts with instrumentation.
% \textbf{Secured(no opt)} are smart contract with instrumentation but
% without optimization. \textbf{Origin} represents the origin smart contracts.

%\subsubsection{Resources Consumption}\label{sec:results:cpu}

Figure~\ref{fig:cpu} shows the resources consumed by Parity for all benchmark
contracts. For each experiment, we measure the performance of the instrumented
smart contracts (\textbf{S}) and the original smart contract (\textbf{O})
respectively. Rows 2-3 present the average CPU usage of Parity. We observe that
the CPU usage of Parity is lower than 10\% for 95\% of time and the average CPU
usage of Parity is lower than 4\% for all benchmarks. Rows 4-5 present the average data written to the
disk per second by Parity.

Our results show that for all contracts, the transaction execution consumes only
a very small portion of the CPU and disk resources. The overhead introduced by
the {\systemname} instrumentation is negligible considering the current
capacity of CPU and disk storage devices and the cost of solving proof of work.
Note that the instrumented ERC1202 VOTE contract consumes sightly more (12 KB/s)
disk write bandwidth because the ERC1202 standard uses a map variable to track
the states for different issues and options, and this map variable is updated
multiple times in \sourcecode{vote} and \sourcecode{createIssue}. This result
validates our previous observation again that the transaction execution is not
the bottle neck of the Ethereum blockchain system. Thus, adding runtime
validation will not downgrade the performance of Ethereum, but improves the
security significantly.

%Besides, the average CPU usage of Parity client is not directly
%related to the complexity of executed smart contracts. This is because the
%block size is limited by gas and the total amount of instructions can be
%executed of each block is limited.

% Row 5-7 presents the total amount of data written to disk by Parity
% client.

% Futhermore, we observe that
% executing secured smart contract requires less CPU resources than
% executing secured smart contract without optimization, which reveals
% that our optimization is effective.

\if 0
\begin{figure*}[t]
    \begin{center}
    \begin{tabular}{ |c|c|c|c|c|c|c|c| }
        \hline
     & & \textbf{BEC-REP} & \textbf{BEC-TRAN}& \textbf{BEC-BAT}
     & \textbf{VOTE} & \textbf{DOZ-REP}& \textbf{DOZ-TRAN}  \\
     \hline
    & \textbf{S}& 7& 3& 7& 5& 17& 14\\
    \cline{2-8}
    \textbf{STORE}& \textbf{S-N}& 7& 3& 7& 15& 18& 15\\
    \cline{2-8}
    & \textbf{O}& 6& 2& 6& 3& 12& 8\\
    \hline
    & \textbf{S}& 9& 5& 9& 7& 10& 11\\
    \cline{2-8}
    \textbf{LOAD}& \textbf{S-N}& 9& 5& 9& 14& 12& 14\\
    \cline{2-8}
    & \textbf{O}& 7& 3& 7& 3& 6& 6\\
    \hline

    \end{tabular}
    \end{center}
    \caption{Number of instructions.}
    \label{fig:instruction}
\end{figure*}
\fi

\subsection{Results without Consensus}

To understand the overhead of the {\systemname} instrumentation under fast
consensus protocols, we run experiments on Parity when we turn off the
consensus layer. We also compare the instrumentation overhead of {\systemname}
with the baseline instrumentation algorithm (which naively performs iterative
checks at the end of each transaction call) to evaluate the effectiveness of
our optimizations.

Figure~\ref{fig:baseline-ERC} presents our experimental results. X axis
corresponds to different smart contracts. The label in the X axis include both
the smart contract name and the TPS of the original contract. Y axis
corresponds to the transaction throughput in the number of transactions
processed by Parity per second (TPS). The Y axis is normalized to the TPS of
the original smart contracts. Red bars in Figure~\ref{fig:baseline-ERC}
correspond to the throughput results of {\systemname}, while blue bars
correspond to the results of the baseline algorithm.

Our results show that even in the extreme cases where the consensus is no
longer the performance bottleneck at all, the instrumentation overhead of
{\systemname} would still be acceptable. The average TPS slowdown caused by the
{\systemname} instrumentation in this set of experiments is 24\%. Our results
also highlight the importance of the delta update and delta check techniques in
{\systemname}. Without those optimizations, the naive instrumentation brings
two orders of magnitude of slowdown in the transaction throughput.

\subsection{Discussion and Future Directions}
\label{sec:results:discussion}

Our experimental results demonstrate that runtime validation is much more
affordable in smart contracts than it is in many other domains. Because of the
performance bottlenecks at the consensus layer and the storage layer,
lightweight runtime instrumentations can have small or even negligible
overhead. Our results reveal several future research directions on how to make
secure and efficient smart contracts and blockchain platforms.

\noindent \textbf{New Languages with Runtime Validation:} To secure smart
contracts, people are designing new smart contract languages that can eliminate
certain classes of errors at compile time, at the cost of limiting
language expressiveness~\cite{move, scilla, obsidian}. Our results imply that making the difficult
trade-off between correctness and expressiveness may not be necessary.
One possible future direction is to design new languages that can better
utilize rigorous runtime validation to enforce the correctness and security.

\noindent \textbf{Static Analysis and Verification:} Developing static analysis
and verification techniques to secure smart contracts is both an active
research topic and an industrial trend~\cite{securify, npchecker, solar, oyente, manticore, mythril,
ethracer, zeus, kevm, verisolid, evmstar, certik, quantstamp}. Like similar techniques in
traditional programs, static analysis techniques often have inaccurate results,
while verification techniques typically require human intervention. Because of
the inexpensive cost, runtime validation can act as backup techniques to cover
scenarios that static analysis techniques fail to fully analyze or that
verification techniques cannot fully prove.

\noindent \textbf{Runtime Validation in Blockchain VM:} {\systemname}
implements runtime validation via Solidity source code instrumentation. We can
further reduce the overhead if we implement some of the instrumented runtime
checks in the blockchain virtual machine, although this would require a hard
fork for existing blockchains like Ethereum.

\noindent \textbf{New Gas Mechanism:} Each EVM instruction charges gas and the
fee of an Ethereum smart contract transaction is determined by the total gas the
transaction consumes. In our experiments, the average gas overhead of the
instrumented contracts is 77.8\%, which is significantly higher than the actual
resource consumption overhead (which is negligible). One possible explanation
is that the gas schedule in Ethereum does not correctly reflect the execution
cost of each EVM instruction. The gas overhead would cause the users of the
instrumented contracts to pay additional transaction fees. In light of our
results, we believe Ethereum and future blockchain systems should adopt a more
flexible gas mechanism to facilitate runtime validation techniques.

\section{Related Work}\label{sec:related}

%We discuss related work in smart contract security and secure compilation.

\noindent\textbf{Smart Contract Security:} There is a rich body of work on
detecting vulnerable smart contracts with different techniques such as symbolic
execution~\cite{maian, solar, oyente, manticore, mythril, ethracer, zeus,verx,
npchecker}, fuzzing~\cite{fuzzing, contract-fuzzing}, domain specific static
analysis~\cite{securify}, and formal verification~\cite{kevm, verisolid,
evmstar, k-verification}. Oyente~\cite{oyente} detect transaction-order
dependency attacks, reentrance attacks, and mishandled exception attacks using
symbolic execution. Verx~\cite{verx} uses delayed abstraction to detect and
verify temporal safety properties automatically. \citeauthor{fuzzing} present a
new fuzzing technique that learns from symbolic execution traces to achieve
both high coverage and high speed~\cite{fuzzing}. Securify presents a domain
specific formal verification technique that translates Solidity smart contract
into Datalog and verifies security properties such as restricted storage writes
and ether transfers~\cite{securify}.

K-framework is a rewrite-based semantic framework that allows developers to
specify semantics of programming language formally~\cite{k-framework}.
KEVM~\cite{kevm} defines the semantic of EVM in $\mathbb{K}$ and verifies the
smart contract against user defined specifications. IELE~\cite{iele} presents a
smart contract virtual machine with a formal specification described in
$\mathbb{K}$ which achieves similar performance as EVM and provides
verifiability.

{\systemname} differs from these previous static analysis, fuzzing, and
symbolic execution approaches in that it inserts runtime checks to enforce user
specified invariants. Unlike these approaches, {\systemname} does not suffer
from false positives and false negatives. Comparing to most formal verification
approaches, {\systemname} is fully automated and does not require human
intervention. Securify is an automated verification tool using SMT solvers, but
it may not scale to complicated contracts due to the potential SMT expression
explosion problem. Also unlike {\systemname} it cannot support sophisticated
constraints like quantifiers. 

%Theses tools formalize the
%semantics of either Solidity language or EVM instructions and
%summarize the behavior of smart contracts or compiled EVM bytecode.
%However due to the nature of static analysis, all of these tools
%suffer from false positives and benign errors.
%Moreover, the time required to analyze a smart contract increases
%exponentially as the complexity of smart contract increases, which
%discourages smart contract developers from using such tools.
%{\systemname} is free from false positives since it rejects
%transactions on the flow based on the current state of the smart
%contract.
%The time for smart contract instrumentation is linear with
%respect to the size of smart contract because {\systemname} simply applies
%a fixed set of rewrite rules to the smart contract.
%Once compiled, the
%constraints inside the standard is enforced and any transaction that
%violates the standard specification will be reverted.

\noindent\textbf{New Language Design:}
Many new programming languages have been proposed recently to improve the smart
contract security. Scilla~\cite{scilla} is a low level smart contract language
with a refined type system which is easy to be verified. Move~\cite{move}
introduces resources types and uses linear logic to enforced access control
policies for digital assets. Obsidian~\cite{obsidian} uses typestate and linear
type to enforce static checks. The trend of these new languages is to sacrifice
expressiveness (e.g., no longer turing-complete) to gain correctness or
security guarantees. Interestingly, our results reveal an alternative path. ---
one possible future direction is to design new languages that can better
utilize rigorous runtime validation to enforce the correctness and security.

%However, none of those languages introduce
%runtime validation. Logic errors such as allowing a user to vote multiple times
%could still be missed by those static checkers.

\noindent\textbf{Secure Compilation:} The security community has been focusing
on secure compilation for a long while. However, such tools are not immediately
applicable to smart contracts because the computation model between EVM and
traditional computer system is quite different. Many existing work focus on
memory safety~\cite{hextype, softboundcets}, side-channel
attacks~\cite{ctwasm}, and error isolation~\cite{isolation}. ContractLarva
presents a runtime verification technique that enforces the execution path of
Ethereum smart contracts to stop vulnerabilities that an attacker invokes a
contract constructor maliciously. Unlike ContractLarva, {\systemname} supports
user defined invariants and works for a much wider scope. 
%To the best of our
%knowledge, {\systemname} is the first secure compilation tool that works on
%Solidity to enforce user defined invariants. 

%, which are rarely considered in
%smart contract security. Besides, none of the above works consider embedding standard
%specification in programs and raising exceptions when the specification
%is violated.

\noindent\textbf{Runtime Checks:} Performing runtime checks is a useful
technique to improve the software security. \citeauthor{control-flow} propose a
framework that enforces the control-flow integrity to mitigate memory
attacks~\cite{control-flow}. \citeauthor{memsafe} enforce both spatial and
temporal memory safety of C programs without introducing high overhead by
modelling temporal errors as spatial errors and removing redundant
checks~\cite{memsafe}. Similarly, Frama-C generates runtime memory monitor
automatically to check E-ACSL specifications~\cite{frama-c}.
\citeauthor{anomaly-detection} describe that patterns in data that do not
conform to expected behavior can help the system to detect intrusions, frauds
and faults~\cite{anomaly-detection}. \citeauthor{diehard}, for example,
implements a runtime system which executes multiple replicas of the same
application simultaneously and detects memory errors dynamically
~\cite{diehard}. {\systemname} inserts runtime checks at the Solidity level.
Modifying the EVM implementation to perform runtime checks may further reduce
the overhead, but it would require a blockchain hard-fork.
%Adopting such tools to detect errors in
%smart contract is not applicable since most of those tools focus on memory
%safety and cannot detect standard violation errors at all.

\section{Conclusion}\label{sec:conclusion}

Runtime validation is an effective and efficient approach to secure smart
contracts. Our results show that because the transaction execution is not the
performance bottleneck in Ethereum, the overhead of runtime validation, which
is often too expensive for other domains, is in fact negligible for smart
contracts.

% {\normalsize \bibliographystyle{IEEEtran}
% }
%% Bibliography
\bibliography{references}
% \printbibliography

\end{document}